\documentclass[aps,prl,twocolumn,groupedaddress, superscriptaddress]{revtex4}

\usepackage{graphicx}
\usepackage{amsmath}
\usepackage{amsthm}
\usepackage{amssymb}

\newcommand{\quantum}{$S=\frac{1}{2}$}

\begin{document}


\title{Spin dynamics in the \quantum\ kagome compound vesignieite,  Cu$_3$Ba(VO$_5$H)$_2$}


\author{R.H. Colman}
\affiliation{Department of Chemistry, University College London, 20 Gordon Street, London, WC1H 0AJ, United Kingdom}
\author{F. Bert}
\affiliation{Laboratoire de Physique des Solides, UMR CNRS 8502, Universit\'e Paris-Sud, 91405 Orsay, France}
\author{D. Boldrin}
\affiliation{Department of Chemistry, University College London, 20 Gordon Street, London, WC1H 0AJ, United Kingdom}
\author{A.D. Hillier}
\affiliation{ISIS facility, STFC, Rutherford Appleton Laboratory,
Chilton, Oxfordshire OX11 0QX, United Kingdom}
\author{L.C. Chapon}
\affiliation{ISIS facility, STFC, Rutherford Appleton Laboratory,
Chilton, Oxfordshire OX11 0QX, United Kingdom}
\author{P. Manuel}
\affiliation{ISIS facility, STFC, Rutherford Appleton Laboratory,
Chilton, Oxfordshire OX11 0QX, United Kingdom}
\author{P. Mendels}
\affiliation{Laboratoire de Physique des Solides, UMR CNRS 8502, Universit\'e Paris-Sud, 91405 Orsay, France}
\author{A.S. Wills}
\email{a.s.wills@ucl.ac.uk}
\affiliation{Department of Chemistry, University College London, 20 Gordon Street, London, WC1H 0AJ, United Kingdom}



\date{\today}

\begin{abstract}
We report the study of high quality samples of the frustrated
\quantum\ kagome antiferromagnet vesignieite, Cu$_3$Ba(VO$_5$H)$_2$.
Neutron powder diffraction measurements evidence the high quality of
the kagome lattice and show no sign of a transition to long-range
order. A kink in the susceptibility below $T=9$~K is matched to a
reduction in paramagnetic-like correlations in the diffraction data
and a slowing of the spin dynamics observed by MuSR. Our results
point to an exotic quantum state below 9~K with coexistance of both dynamical and small frozen moments
$\sim 0.1 \mu_B$. We propose that
Dzyaloshinsky-Moriya interaction is large enough in this system to
stabilize this novel quantum ground state.

\end{abstract}

\pacs{PACS numbers: }

\maketitle


The low-connectivity and dimensionality of a kagome lattice of
antiferromagnetically coupled \quantum\ ions (KAFM) make it  the
ideal framework to study one of the most sought after states, the
quantum spin liquid (QSL) \cite{kagomeQSL, Claire, Mila, Hastings}.
Despite the advances in theoretical understanding of QSLs, progress
in the field is hindered by a lack of physical examples of \quantum\
KAFMs. Even the best model systems possess noteworthy shortcomings,
such as structural distortions, disorder and further neighbor
exchange.

Much of the interest in KAFMs was inspired by evidence of a
liquid-like ground state in the kagome-like compound
SrCr$_{9\text{p}}$Ga$_{12-9\text{p}}$O$_{19}$ (SCGO) where persistant spin-dynamics were observed down to $100$~mK using muon
spin relaxation spectroscopy (MuSR), far below a spin-glass transition at $T_{\text{g}}
\simeq 3.5$~K \cite{UemuraSCGO}.
Following on from this work, volborthite, Cu$_3$V$_2$O$_7$(OH)$_2\cdot 2$(H$_2$O),
was believed for several years to be the closest approximation to a
\quantum\ KAFM. Susceptibility measurements showed it to have strong
antiferromagnetically coupled spins ($\theta_{\text{W}}=-115$~K)
\cite{HiroiVolborthite} that remain dynamic as $T\rightarrow
0$~\cite{Fukaya}, despite partial spin freezing
\cite{MendelsVolborthite}. This lack of N\'eel order is remarkable
given the monoclinic distortion to the lattice: the triangles of
volborthite are isosceles and the inequivalence of the Cu\---O\---Cu
superexchange pathways is expected to reduce the ground state
degeneracy.

Undistorted kagome lattices of \quantum\ ions have more recently
been studied in several closely related materials: Zn- and
Mg-herbertsmithites, kapellasite and haydeeite
\cite{NoceraHerbertsmithite, DeVriesHerbertsmithite,
ColmanMgHerbertsmithite, NoceraMgParatacamite, ColmanHaydeeite,
ColmanKapellasite}. These systems display the characteristic  properties of frustrated magnets, such as suppression of
magnetic
ordering~\cite{MendelsHerbertsmithiteMuSR,Helton,ColmanKapellasite}
and have revealed new effects such as temperature-independent
magnetic excitation spectra \cite{DeVriesInelastic}. They all also,
however, feature some degree of dilution of the moment-bearing
Cu$^{2+}$ ions by diamagnetic metal ions (Zn$^{2+}$/Mg$^{2+}$) that
complicates understanding of the ground state properties
\cite{ColmanHaydeeite, MendelsO17, BertHerbertsmithite,
DeVriesHerbertsmithite}.

In order to differentiate the intrinsic properties of the KAFM from
the effects caused by deviations from ideality, new model materials
close to the idealized \quantum\ kagome-structure are needed. One
recently proposed example is the mineral vesignieite,
Cu$_3$Ba(VO$_5$H)$_2$ \cite{HiroiVesignieite, VesignieiteESR,
LafontainVesignieite}. Unlike the herbertsmithites, the kagome
structure of vesignieite is not created by diamagnetic dilution of
another lattice, and so there is little possibility of
anion-exchange leading to structural defects within the KAFM.

In this letter we investigate high quality samples of vesignieite
 by powder neutron diffraction, magnetic
susceptibility and $\mu$SR spectroscopy. We show that the kagome
lattice is very close to the ideal geometry and therefore that
vesignieite provides an opportunity for the study of minimally
perturbed KAFM physics.

\begin{figure}[ht]
\begin{center}
    \includegraphics[width=0.5\textwidth]{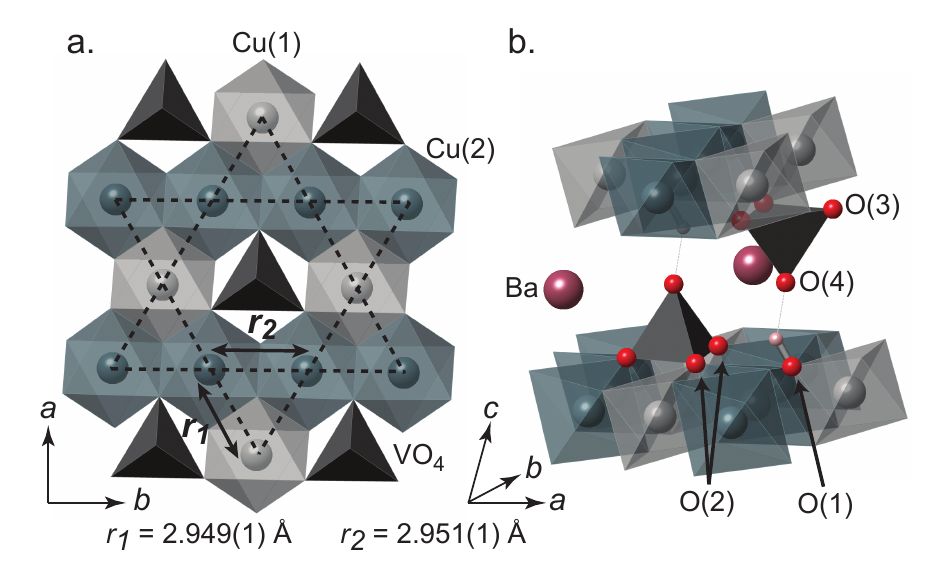}
\end{center}
\caption{The kagome lattice of vesignieite has two crystallographically distinct CuO$_6$ octahedra, and near 3-fold rotational symmetry. The kagome planes are coupled by VO$_4$ tetrahedra, hydrogen bonded to hydroxyl groups in the neighbouring planes and interstitial Ba$^{2+}$ ions. There are 4 types of oxygen positions: O1, the hydrogen bonding O\---H group; O2 and O3, define the basal plane of the VO$_4$ tetrahedra; and O4 is the tetrahedral apex.}
\label{Fig1:Structure}
\end{figure}


No structural refinements were presented of the previously synthesized samples of vesignieite~\cite{HiroiVesignieite}. Inspection of the
broad powder X-ray diffraction (XRD) peaks, however,  indicates that they are poorly crystalline, with
coherent order restricted to length scales as low as a few nm. Such deficiency will disrupt the
magnetic behavior of the KAFM and complicate its interpretation.


We used an alternative synthetic method, based on the evaporation of
ammonia from a solution of Schweizer's reagent \cite{AmmoniaPrep},
[Cu(NH$_3$)$_4$(H$_2$O)$_2$](OH)$_2$ (0.678~g Cu(OH)$_2$ in 100~ml
28\,\% NH$_4$OH solution), vanadium pentoxide, V$_2$O$_5$ (0.154~g)
and barium acetate, Ba(CHOO)$_2$ (0.217~g), heated to reflux for
$\sim$4~hrs. The evaporation of ammonia causes the slow release of
Cu(OH)$_2$ into solution and the controlled crystallization of
vesignieite. To further improve crystallinity, the vesignieite
sample ($\sim$1~g) was then hydrothermally annealed in (25~ml)
H$_2$O (or 99.9 atom~\% D$_2$O for the deuterated sample), at
190~$^\circ$C, for 48~hrs. The resultant vesignieite sample was
found to be phase pure by powder XRD.

\begin{figure}[ht]
\begin{center}
    \includegraphics[width=0.4\textwidth]{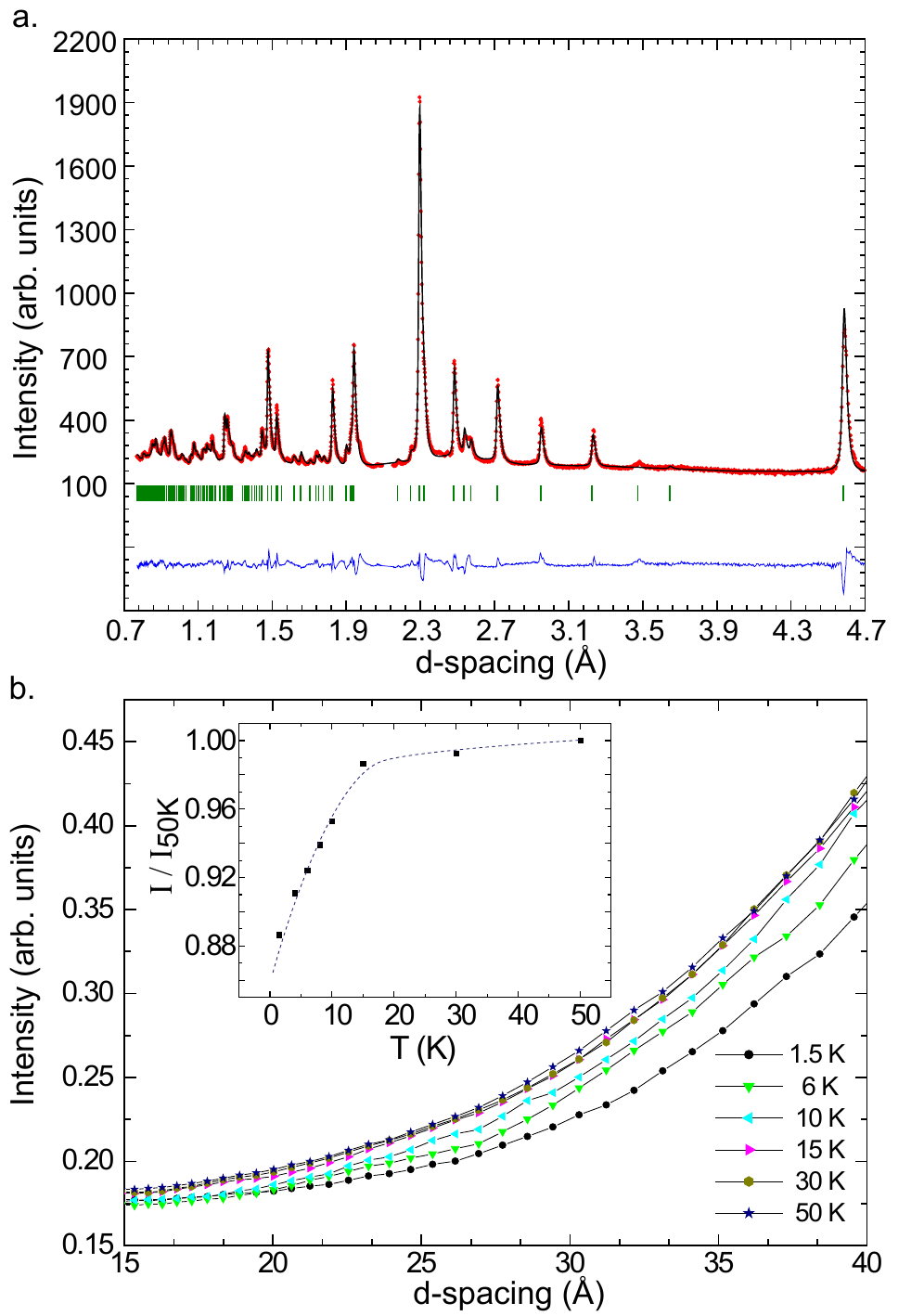}
\end{center}
\caption{a. PND data collected from a deuterated vesignieite sample
at $T=1.5$~K. The red points are the data, the black line is the
fit, the blue curve is the difference and the tick marks indicate
reflection positions. Refined lattice parameters, in the monoclinic
spacegroup C~$1\,2/m\,1$ (\# 12) are $a=10.210(2)$~\AA,
$b=5.903(1)$~\AA, $c=7.739(1)$\AA \ and $\beta =116.15(1)^\circ$.
The final R$_{\text{WP}}$ is 3.61. b. Plot of the high d-spacing
data at several temperatures. Inset: integrated intensity of the
scattering in the range $15<d<40$~\AA\ as a function of temperature,
normalised to the 50~K scattering. The dashed line is a guide to the
eye.} \label{Fig2:Refinement}
\end{figure}

Powder neutron diffraction (PND) data were collected between 1.5 and
50~K on a deuterated sample of vesignieite, Cu$_3$Ba(VO$_5$D)$_2$,
(1.75~g), using the  diffractometer WISH (ISIS, UK). Rietveld
refinements of the crystal structure of vesignieite afforded the fit
shown in Fig.~\ref{Fig2:Refinement}. Our refined parameters show
that the isosceles triangle built from the Cu(1) and Cu(2) sites is
very close to equilateral with $r_1=2.949(1)$\AA \ and
$r_2=2.951(1)$\AA . This distortion of 0.07~\% is far less than that
of previous samples \cite{HiroiVesignieite} and greatly strengthens
the case for vesignieite as an important model system with which to
probe \quantum\ kagome physics.

Over the range 1.5~K - 50~K, no change was seen in the Bragg
diffraction other than a small thermal lattice-contraction; there
was no evidence of magnetic-Bragg diffraction, or broad features
indicative of short-range spin correlations. Remarkably, a reduction
in the form factor-type scattering at high d-spacing is observed
below T $\sim 10$~K, indicating that below this temperature the
paramagnetic-like component is being reduced though no other
magnetic scattering is seen to develop
(Fig.~\ref{Fig2:Refinement}b).

\begin{figure}[!ht]
\begin{center}
    \includegraphics[width=0.45\textwidth]{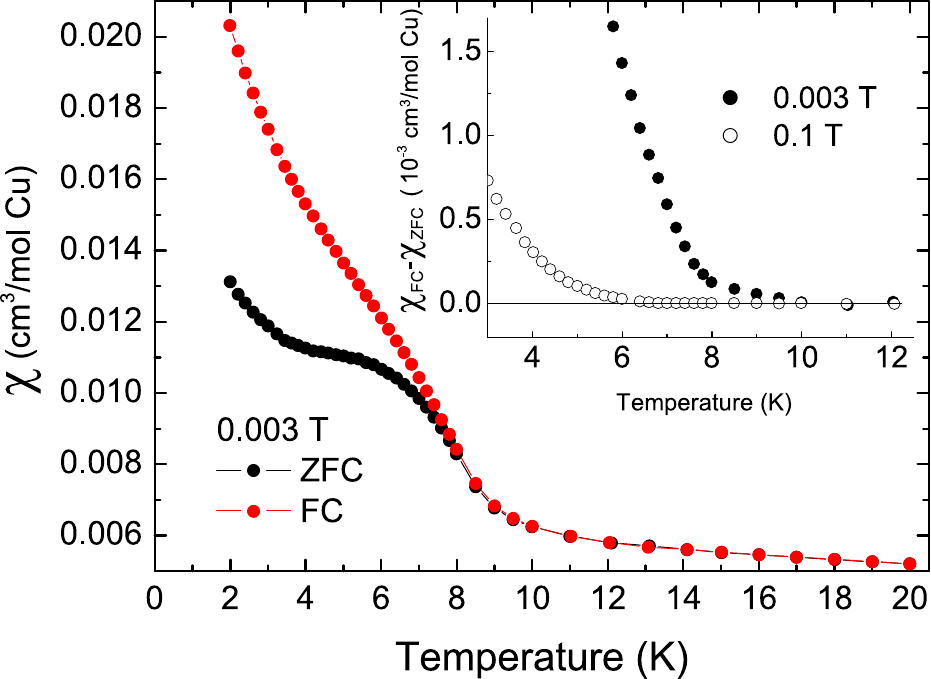}
\end{center}
\caption{a. The temperature-dependent zero-field-cooled (ZFC) and
field-cooled (FC) DC-susceptibility measured in a field of 0.003~T.
Inset: difference of the ZFC and FC susceptibilities.}
\label{Fig3:Susceptibility}
\end{figure}

DC-magnetic susceptibility measurements were performed using an MPMS
SQUID magnetometer. In the region $100<T<300$~K, for applied field
$\mu_0H>1$~T, the data displays a typical Curie-Weiss response that
could be fitted to give a Weiss temperature
$\theta_{\text{W}}=-85(5)$~K, indicative of antiferromagnetic
exchange, and an effective moment
$\mu_{\text{eff}}=2.02(2)$\,$\mu_B$ in agreement with published
data~\cite{HiroiVesignieite}. As in Ref.~\cite{HiroiVesignieite}, a
kink in the susceptibility is observed at $T\sim 9$~K. Our data show that this feature  is
also accompanied by an opening of the field cooled and zero field
cooled branches (Fig.~\ref{Fig3:Susceptibility}), which is most
obvious at low field, and is sugggestive of spin glass-like freezing.

As a local probe, $\mu$SR is a technique of choice to discriminate
between extrinsic and intrinsic magnetic contributions. It has
proven to be a powerful tool for probing spin dynamics in frustrated
materials and in the search for spin liquid ground states. A
protonated sample of vesignieite was investigated using the
spectrometer MUSR (ISIS, UK). As muons, $\mu^+$, implant near
centers of negative charge, the 4 crystallographically distinct oxygen
positions in vesignieite (O1\---O4, Fig.~\ref{Fig1:Structure}) give
rise to 4 different likely stopping positions.

\begin{figure}[ht]
\begin{center}
    \includegraphics[width=0.45\textwidth]{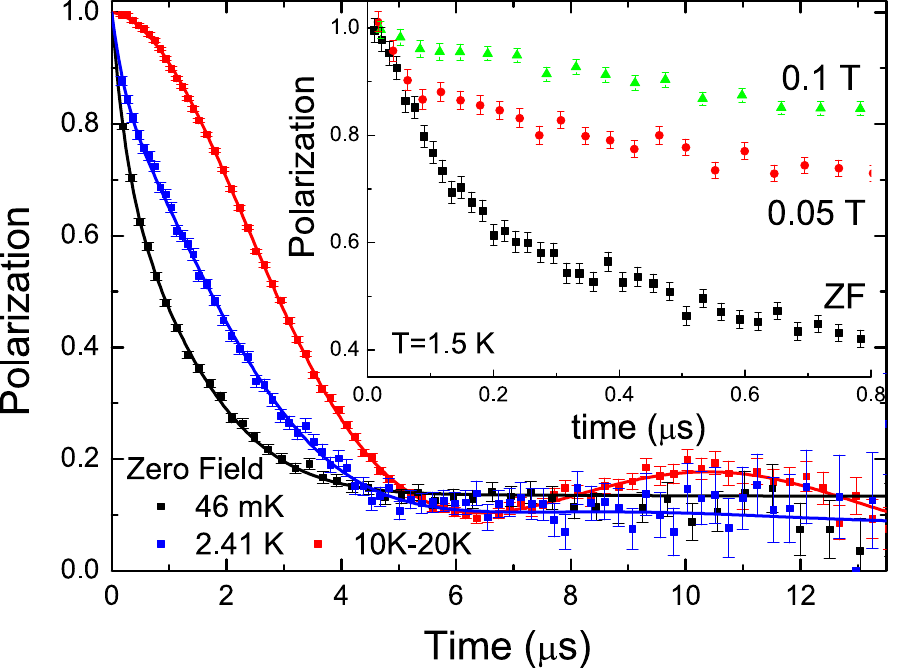}
\end{center}
\caption{Muon decay asymmetry versus time in zero applied field.
Inset: initial asymmetry at 1.5~K in zero and longitudinal field.}
\label{MuSRfig1}
\end{figure}

For $T>9$~K, the muon relaxation is dominated by the nuclear magnetism of
Cu, V and H nuclei, which is static at the muon time scale. Muons
also experience the Cu$^{2+}$ electronic spin-dynamics, modeled by an
additional slow exponential decay. For required simplicity, the same relaxation rate, $\lambda$, for all muon sites was imposed. The zero field asymmetries (Fig.~\ref{MuSRfig1}) were therefore fitted to
$P_{\text{para}}(t)=P_{\text{nuc}}(t)e^{-\lambda t}$ where the
temperature independent $P_{\text{nuc}}$ is the static
nuclear contribution~\cite{nuclear_part}.
 Above $T=15$~K,
 an almost temperature independent
rate $\lambda=0.012(1)$~$\mu$s$^{-1}$ was seen, which arises from exchange
fluctuations of the Cu$^{2+}$ electronic spins with the rate $\nu
\simeq J/2\hbar$~\cite{UemuraSCGO}. Using
$J/k_B=53$~K~\cite{HiroiVesignieite}, the relaxation in the
paramagnetic limit $\lambda=2\Delta H / \nu$ yields the width
$\Delta H \simeq 0.15$~T for the random field distribution at the
muon sites. This value is in fair agreement with the crude estimate
of the average dipolar field $\simeq 0.12$~T arising from the
fluctuating $\simeq 1 \mu_B$ copper moments and experienced by muons
located close to the oxygen sites.

Below $T\sim9$~K  the relaxation of the asymmetry steeply increases,
a change that cannot simply be accounted for by a slowing down of
the electronic spin dynamics. In addition, a fast initial relaxation
develops, and the polarisation as $t\rightarrow \infty$ is no longer
0. These are the usual signatures of the presence of a magnetic
phase, static on the muon time scale. To gain more insight into the
nature of this frozen phase, spectra were recorded at $T=1.5$~K on
the GPS instrument at the Paul Scherrer Institut facility (CH),
which provides high resolution at short times (see inset of
Fig.~\ref{MuSRfig1}). The absence of spontaneous oscillations in the
zero field muon polarization decay points to a highly disordered
magnetic state of the spin-glass type, which is consistent with the
absence of magnetic Bragg peaks in PND. Moreover a field applied
along the initial muon polarization direction, of $\sim0.1$~T is
required to overcome the internal static field-distribution. This
indicates that the maximal static internal field at the muon sites
is $\sim 0.01$~T. This value is in contrast to the former estimation
of the dipolar field $\sim$0.12~T generated by full 1~$\mu_B$ copper
moments. We deduce that the frozen moments in the static phase are
surprisingly small, of the order of 0.1~$\mu_B$. This suggests that
not all the spin degrees of freedom are freezing below 9~K or that
strong quantum fluctuations are present in the frozen phase.

\begin{figure}[t!]
\begin{center}
    \includegraphics[width=0.4\textwidth]{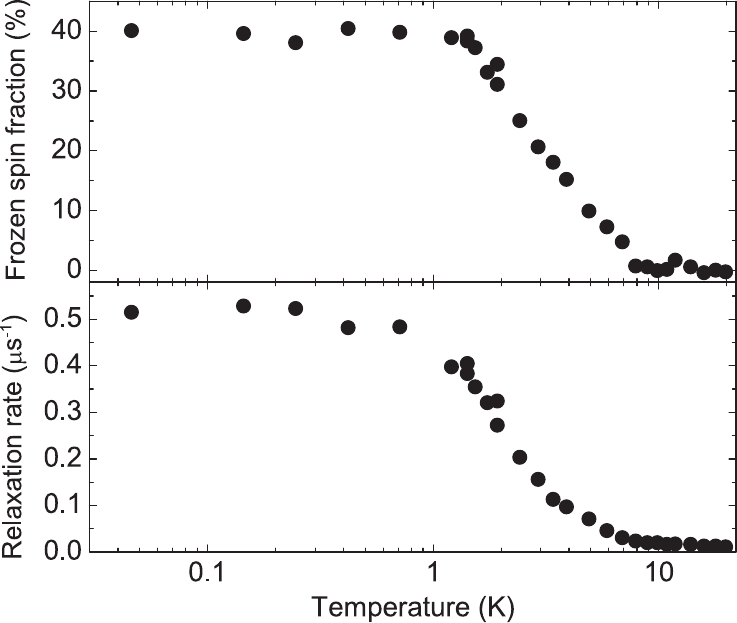}
\end{center}
\caption{Temperature dependence of the frozen spin fraction
$f_{\text{f}}$ and the muon relaxation rate $\lambda$ determined
from the fits of the zero-field asymmetries (see text).}
\label{MuSRfig2}
\end{figure}

The muon asymmetry across the whole temperature range could be
fitted with the following function
$P(t)=f_\text{f}P_{\text{f}}(t)+(1-f_{\text{f}})P_{\text{para}}(t) $, where
$f_{\text{f}}$ monitors the fraction of spins involved in the frozen
magnetic phase. Given the limited accuracy at short time of the ISIS
data, the minimal, temperature independent form
$P_{\text{f}}(t)=\frac{2}{3}e^{-t\lambda_{\text{fast}}}+\frac{1}{3}
$, where $\lambda_{\text{fast}}=4(1)\, \mu\textrm{s}^{-1}$ was used to
account for the static spin fraction. The frozen fraction,
$f_{\text{f}}$, is found to increase gradually below $T=9$~K at the
expense of the paramagnetic, non-frozen, fraction (see
Fig.~\ref{MuSRfig2}a) down to $T\sim1$~K where it saturates at
$\sim40$~\% of the sample spins. Simultaneously, the relaxation in
the non-frozen phase gradually increases on cooling below $T=9$~K
and eventually saturates at $\lambda=0.50$(2) $\mu$s$^{-1}$ below
1~K (see Fig. \ref{MuSRfig2}b) indicating persistent, slow dynamics
as $T \rightarrow 0$. Although the origin of this behavior remains
unclear, it is reminiscent of many other frustrated
antiferromagnetic systems such as kagome bi-layers
\cite{UemuraSCGO}, volborthite \cite{Fukaya} or even rare earth pyrochlores \cite{PyrochloreMuSR} and points to an exotic
quantum ground state in the unfrozen spin fraction of the sample.
The similar temperature evolutions of the frozen fraction,
$f_{\text{f}}$, and the relaxation rate, $\lambda$, in the
paramagnetic phase indicates that these two responses are within the
same framework and neither is an extrinsic contribution. Our $\mu$SR
results clearly demonstrate the slowing down of all the electronic
spins below 9~K. This rules out the previous interpretation of the 9~K
upturn in the susceptibility, as arising from 7\% free impurity
spins \cite{HiroiVesignieite}. The 9~K feature in the $\mu$SR data instead points to the building up of
intrinsic spin-spin correlations in vesignieite.

In summary, structural refinements of PND data show high quality
samples of vesignieite to have near 3-fold rotation symmetry of the idealized
magnetic kagome lattice. Susceptibility, diffraction data and
$\mu$SR data consistently evidence a strong slowing of the spin
dynamics below 9~K, down to weak static magnetism for $\sim 40\%$
 of the Cu spins. It is enlightening to compare the exotic
ground state properties of vesignieite to other well studied quantum
kagome materials. In structurally perfect herbertsmithite, where all
Cu$^{2+}$ ions are identical by three fold symmetry, no transition
to a static state, even of a small spin fraction, is detected down
to at least $J/4000$~\cite{MendelsHerbertsmithiteMuSR}. In
volborthite, the kagome lattice is sizeably distorted $\sim
3\%$~\cite{HiroiVolborthite,HiroiVesignieite}. Partial static
magnetism was detected in NMR and SQUID measurements below $\sim
J/80$ whereas $\mu$SR relaxation is dominated by coexisting spin
dynamics. In this context, it is surprising to observe a spin-state
reminiscent of that in volborthite in the quasi-perfect kagome
material vesignieite at a high temperature of $\sim J/6$. At the
classical level, distortion of the kagome lattice hinders the
connection of the degenerated ground states by local zero-energy
modes and may favor spin glass physics~\cite{Wang}. It seems,
however, unlikely that the minute distortion of $\sim 0.07\%$ in
vesignieite is responsible for partial freezing at $\sim J/6$.
Dzyaloshinsky-Moriya interaction, which is allowed to occur on the
kagome lattice, is another possible perturbation to the pure KAFM.
It may destabilize the kagome spin liquid ground state if  its
magnitude, $D$, is $D/J \gtrsim 0.1$~\cite{Cepas} . Analysis of the
ESR line width has provided the value $D/J \sim 0.08$ for
herbertsmithite~\cite{Zorko}. Although a detailed understanding of
the ESR lineshape is quite involved, we may assume a similar origin
of the room temperature width $\Delta H \sim D^2/J$ also in
vesignieite.  Given that $\Delta H$ is close to the value seen in
herbertsmithite \cite{VesignieiteESR}, whilst $J$ is three times
smaller, we calculate that $D/J$ would be $\sim 0.14$, {\it i.e.} on
the N\'eel ordered side of the predicted critical point
\cite{Zorko}. Further studies are needed to ascertain the exact
ratio of $D/J$ and to understand why long range order is not
stabilized despite its apparently large value, and why a dynamic
liquid character persists for a majority spin fraction as $T
\rightarrow 0$. Being a very clean system, vesignieite is certainly
a good material to address experimentally these pending questions
related to the criticality of KAFM physics.

\begin{acknowledgments}
We acknowledge technical assistance of A. Amato in $\mu$SR
measurements at PSI and enlightening discussion with A.~Zorko. The
work was partly supported by the ANR-09-JCJC-0093-01 grant and EC FP
6 program, Contract No. RII3-CT-2003-505925.
\end{acknowledgments}


\begin{thebibliography}{20}
\bibitem{nuclear_part}
We used  $P_{\text{nuc}}(t)=
f_1P_{\text{OH}}(\omega,\Delta_{\text{OH}},t)+f_{23}P_{\text{KT}}(t,\Delta
H_{23}) +f_4P_{\text{KT}}(t,\Delta H_4)$ where $f_1$, $f_4$ and
$f_{23}$ are the fractions of muons stopping close to O1, O4, and to
both O2 and O3 which feature extremely similar crystallographic
environments. $P_{\text{KT}}(t,\Delta H)$ is the familiar
Kubo-Toyabe relaxation function expected for a static field
distribution of width $\Delta H$. Muons stopping next to O1 form a
$\mu$\---OH complex which yields a specific oscillating
relaxation~\cite{MendelsHerbertsmithiteMuSR,Brewer86}. The following
values were refined: $f_1 \sim f_{23} \sim f_4=0.33(5)$, $\Delta
H_{23}=1.7(1)$~G, $\Delta H_4=4.2(1)$~G,
$\omega/\gamma_\nu=4.6(1)$~G and $\Delta_{\text{OH}}=0.7(3)$~G.
\end{thebibliography}

\end{document}